\documentclass[letterpaper]{article}
\usepackage{aaai2026}
\usepackage{times}
\usepackage{helvet}
\usepackage{courier}
\usepackage[hyphens]{url}
\usepackage{graphicx}
\usepackage{natbib}
\usepackage{quoting}
\usepackage{caption}
\usepackage[inline]{enumitem}
\quotingsetup{leftmargin=10pt, rightmargin=10pt}

\newcommand{\ie}{\emph{i.e.}}
\newcommand{\eg}{\emph{e.g.}}

\newcommand{\vs}{\emph{vs. }}

\title{Information Access of the Oppressed:\\Freirean Design for Emancipatory Information Access}
\author{
Bhaskar Mitra\textsuperscript{\rm 1},
Nicola Neophytou\textsuperscript{\rm 1},
Sireesh Gururaja\textsuperscript{\rm 2}
}
\affiliations {
\textsuperscript{\rm 1}Independent Researcher, Tiohtià:ke / Mooniyang / Montréal, Canada\\
\textsuperscript{\rm 2}Carnegie Mellon University, Pittsburgh, USA\\
bhaskar.mitra@acm.org,
neophytounicola@gmail.com,
sgururaj@cs.cmu.edu
}

\begin{document}
\maketitle

\begin{abstract}
Online information access (IA) platforms are targets of authoritarian capture.
We explore the question of how to safeguard our platforms and ensure emancipatory outcomes through the lens of Paulo Freire's theories of emancipatory pedagogy.
Freire's theories provide a radically different lens for exploring IA's sociotechnical concerns relative to the current dominating frames of fairness, accountability, and transparency.
We make explicit, with the intention to challenge, the \textit{technologist-user dichotomy} in IA platform development that mirrors the teacher-student relation in Freire's analysis.
By extending Freire's analysis to IA, we critique the \textit{technologists-as-liberator} frame where it is the burden of (altruistic) technologists to mitigate the risks of emerging technologies for marginalized communities.
Instead, we advocate for \textit{Freirean Design} whose goal is to structurally expose the platform for co-option and co-construction by community members in aid of their emancipatory struggles.
\end{abstract}
\section{Introduction}
\label{sec:intro}
On July 23, 2025, U.S. president Donald J. Trump issued an executive order titled ``Preventing Woke AI in the Federal Government''~\cite{trump2023preventing} prohibiting federal procurement of AI models that incorporate concepts like ``critical race theory, transgenderism, unconscious bias, intersectionality, and systemic racism; and discrimination on the basis of race or sex''.
The executive order, while claiming to safeguard AI models from alleged ``ideological biases'', seeks to strong-arm AI companies to conform with the Trump regime’s ideological agenda~\cite{noble2025president}.
Likewise, China's National Technical Committee 260 on Cybersecurity of SAC published a technical document\footnote{\url{https://www.tc260.org.cn/upload/2024-03-01/1709282398070082466.pdf}} mandating all Chinese generative AI models to not generate content that violates the country’s ``core socialist values''.
Consequently, the Chinese AI chatbot DeepSeek refuses to answer questions on certain political topics, like Tiananmen Square and Taiwanese independence~\cite{lu2025we}.
Such acts of ideological imposition on AI model outputs is not limited only to state actors.
In May 2025, xAI's chatbot Grok promoted conspiracy theories about ``white genocide'' in South Africa~\cite{kerr2025musk} after said conspiracy theories were mainstreamed by xAI's owner Elon Musk~\cite{imray2025south}.
On another occasion the same year, Grok responded to users with antisemitic and pro-Nazi content~\cite{taylor2025musk}, which followed only a few months after Musk himself was embroiled in a controversy for making hand gestures at Trump's presidential inauguration event that appeared to be Nazi salutes~\cite{jacobsen2025ein}.

As AI becomes widely used for information access (IA)\footnote{Information access  refers to the focused interaction between people and information~\cite{shah2024envisioning}. Subfields of the study of IA include information retrieval (IR) and information filtering (IF). Search engines, recommender systems, and chatbots are common examples of IA systems. While social media may also facilitate access to information, their primary function is to support interactions between people and therefore are not IA systems.}~\cite{lardinois2023microsoft, dastin2023google, chatterji2025people}, these instances of attempted capture emulate historical examples of regimes trying to manipulate different forms of media~\cite{dragomir2025old}, and fit into a growing trend of \emph{digital authoritarianism}~\cite{roberts2025digital}---defined as ``the use of digital information technology by authoritarian regimes to surveil, repress, and manipulate domestic and foreign populations''~\cite{polyakova2019exporting}.
Traditional media, social media, and IA platforms have historically been sites of conflict between oppression and emancipation because they shape public beliefs and opinions, for example, how search results can influence voting preferences~\cite{epstein2015SEME}.
However, these concerns are particularly serious today in light of:
\begin{enumerate*}[label=(\roman*)]
    \item Rising levels of democratic erosion~\cite{gorokhovskaia2024mounting} and geopolitical conflicts~\cite{taylor2023historic, un2023highest},
    \item growing concentration of economic and political power in the hands of Big Tech~\cite{oremus2017big, whittaker2021steep, wef2024world, KakWest2023} which increasingly aligns itself with authoritarian actors~\cite{lafrance2024rise, becca2025headed, kanj2025big, fatafta2024big, akbari2025big, kang2024silicon, kang2025big}, and
    \item emerging capabilities of generative AI, such as \emph{AI persuasion}~\cite{burtell2023artificial, carroll2023characterizing, park2023ai, el2024mechanism} that maybe weaponized within IA platforms for mass manipulation~\cite{mitra2024sociotechnical}.
\end{enumerate*}
As \citet{mitra2025emancipatory} warns: ``\emph{Imagine every time you searched online or accessed information via your digital assistant, the information was presented to you exactly in the form most likely to alter your consumer preferences or political opinions.
Or, consider AI-generated digital characters in ads and videos that appropriate marginalized identities to say or act in ways that real members of that community may be strongly opposed to---a new form of `Digital Blackface'~\cite{johnson2023generative}.}''
The confluence of these factors calls for more scrutiny on the evolving relationship between users and IA platforms.

While AI capabilities may make IA platforms more attractive for capture, the risk of capture is salient to IA platforms more generally.
Algorithmic or policy interventions for AI technologies alone are unlikely to be sufficient; and safeguarding AI models from oppressive ideological imposition by their owners is inherently difficult given the lack of transparency around how these models are trained by private corporations.
Instead, these risks are better analyzed and addressed in the context of the IA platforms where these models are deployed and in the context of their complex interactions with people and information.
Centering the IA platform also crucially leaves room for refusal to deploy these models as a valid form of resistance~\cite{barocas2020not}.
Therefore, we ask: ``\emph{What is the alternative IA infrastructure that we should reimagine and realize in practice to mitigate authoritarian capture of IA platforms?}''.

Our pursuit for alternative IA infrastructure is not value-neutral.
Our motivation stems from the recognition of IA as sites of political struggle between oppression and emancipation.
We reject any ambivalence or false neutrality with respect to oppressor-oppressed relations and situate our work in recent calls to explicitly align IA with humanistic and emancipatory goals~\cite{mitra2025emancipatory, trippas2025report}.
It is our emancipatory aspirations that prompt us to recognize the grave risks of state and corporate actors capturing our information platforms to manipulate mass opinions.
This is in contrast to other frames of critique of platform capture, like \emph{digital sovereignty}~\cite{djuric2025canada, fitzGerald2025digital} that do not directly target the practices of capture, but is fueled by discontent about whose nationalist interests are served by platforms.
In contrast, we are concerned by how oppressive power relations subverts knowledge production and consensus building, suppresses marginalized voices, and perpetuates epistemic injustices~\cite{fricker2007epistemic}.
This distinction has practical implications for our work.
For example, in the social media landscape, decentralization has emerged as a dominant strategy to safeguard against capture~\cite{ciriello2025decentralized}, which may also be key to protecting IA platforms.
However, the decentralization of technological infrastructure alone is insufficient to ensure that platforms are emancipatory as they tend to reproduce dominant structures of oppression~\cite{hendrix2022whiteness}.
Instead, we need a broader reenvisioning of IA platforms.

Practices of futuring~\cite{fry2009design} and envisioning~\cite{reeves2012envisioning} shape technology development.
Our sociotechnical imaginaries~\cite{jasanoff2015dreamscapes} do not only imagine but also co-construct our digital futures~\cite{mager2021future}.
In IA, these practices commonly manifest as scholarly publications---\eg, \cite{shah2024envisioning, metzler2021rethinking}---and strategic workshops---\eg, \cite{trippas2025report, azzopardi2024search, hagen2026retrieval}.
However, the dominant values and epistemologies that shape these visions often remain implicit~\cite{vrijenhoek2023report, mitra2025search}, and insulated from community purview.
Here, envisioning emancipatory IA poses two distinct requirements.
The first requirement is that the envisioning process should produce a figurative blueprint for IA platforms that safeguards against authoritarian capture and contribute towards emancipatory societal outcomes.
A second less obvious but nonetheless critical requirement is that the envisioning process should involve reflective participation of marginalized communities, who should be integral to shaping the values of such platforms.
We motivate this second requirement by drawing from critical scholarship on emancipatory pedagogy, in particular Paulo Freire's canonical work ``\emph{Pedagogy of the Oppressed}''~\cite{freire1970pedagogy}.
In the context of emancipatory pedagogy, Freire emphasizes that the oppressed are not simply targets but rather the executors of liberatory actions.
He critiques what he refers to as the ``\emph{banking concept of education}'' in which emancipatory knowledge is ``gifted'' by those committing the act of liberating to those who are purportedly being liberated.
He writes:
\begin{quoting}
  ``\emph{Attempting to liberate the oppressed without their reflective participation in the act of liberation is to treat them as objects which must be saved from a burning building; it is to lead them into the populist pitfall and transform them into masses which can be manipulated~[\ldots]~The conviction of the oppressed that they must fight for their liberation is not a gift bestowed by the revolutionary leadership, but the result of their own conscientização.}'' (Pages 65-67)
\end{quoting}

When we embark to envision emancipatory IA platforms, we should vigorously reject the notion that it is the burden of select individuals or institutions to determine the design that makes platforms emancipatory.
We should also reject \emph{technosolutionism} and notions that there exists key elusive design tricks waiting to be discovered that once incorporated render IA platforms universally emancipatory and safe from capture in all diverse cultural and sociopolitical contexts.
Rather, the key to emancipatory IA lies in holding space for reflective participation of community members in the process of envisioning, design, implementation, maintenance, and governance of platforms.
We operationalize this insight by conceptualizing \textit{Freirean design} that commits to engage historically marginalized communities in the active co-option and co-construction of technology.
In Freirean design, the platform is intentionally underdesigned to make room for diverse reflection, dialogue, participation, collaboration, and negotiation by community members in the envisioning and actualization of emancipatory IA.
We foreground community participation to forgo the modes of technology production where people relate to technology either as technologists (\emph{who envision, build, and maintain the platform}) or users (\emph{who use but cannot change the platform}); in favor of futures where we are all engaged in epistemic activities of knowledge production and consensus building through the making and unmaking of IA technologies.

The rest of this article is organized as follows: We situate our work in the existing traditions of justice-oriented computing in Section~\ref{sec:background}.
Next, we develop the concept of Freirean design drawing from his theories and analyses in Section~\ref{sec:freire}.
In Section~\ref{sec:problem}, we discuss how to put this design methodology into practice using Freire's problem-posing approach.
Finally, in Section~\ref{sec:discussion}, we reflect on the salient tensions involved in this work before concluding in Section~\ref{sec:conclusion}.

\section{Justice-oriented computing}
\label{sec:background}

Several subfields of computing, like digital humanism, science and technology studies (STS), and human-computer interaction (HCI) have studied how computing intersects with concerns of justice and emancipation and how to reorient the field to support humanistic values and affect social justice.
The Vienna manifesto on Digital humanism~\cite{werthner2020vienna} demands that ``we must shape technologies in accordance with human values and needs, instead of allowing technologies to shape humans''; STS goes beyond studying the ``origins, dynamics, and consequences of science and technology'' to engage with questions of equity, politics, and social change~\cite{barben2008handbook}; and HCI has cultivated a tradition of social justice research~\cite{dombrowski2016social, chordia2024social} and design activism~\cite{thorpe2011defining, fuad2013design, hauser2013design} within its discipline.
Salient to their concerns, is the question of what ``better'' futures we want to build for and the values that shape our future imaginaries.
Approaches such as value-sensitive design~\cite{friedman1996value} and speculative design~\cite{auger2013speculative} foreground these questions but subsequently leave open the critical enquiries: for \textit{what} values~\cite{albrechtslund2007ethics, manders2011values, reijers2019moving, jacobs2021value} and ``better'' according to \textit{who}~\cite{bowen2010critical, martins2014privilege}?

\subsection{Humanistic and emancipatory values in computing}
\label{sec:background-emancipatory}
To address the ``value'' question in computing, \citet{asad2019prefigurative} proposed prefigurative design to explicitly reorient design work in service of ``political or civic goals'' and several other strands of research has explicitly grounded their work in liberatory epistemologies, including humanistic~\cite{bardzell2016humanistic, bardzell2015humanistic}, anti-oppressive and emancipatory~\cite{smyth2014anti, bardzell2016humanistic, kane2021avoiding, monroe2021emancipatory, saxena2023artificial}, feminist~\cite{wajcman2004technofeminism, wajcman2010feminist, bardzell2010feminist, bardzell2011towards, bardzell2011feminism, bardzell2016feminist, bardzell2018utopias, d2020data}, queer~\cite{light2011hci, klipphahn2024introduction, guyan2022queer}, postcolonial and decolonial~\cite{irani2010postcolonial, philip2012postcolonial, dourish2012ubicomp, sun2013critical, ali2014towards, ali2016brief, akama2016speculative, irani2016stories, avle2017methods, adams2021can, mohamed2020decolonial}, anti-racist~\cite{abebe2022anti, ogbonnaya2020critical}, indigenous~\cite{winschiers2013toward, shaw2014mobile, akama2016speculative}, anti-casteist~\cite{kalyanakrishnan2018opportunities, sambasivan2021re, vaghela2022interrupting, vaghela2022caste, shubham2022caste, kanjilal2023digital}, anti-ableist~\cite{williams2021articulations, sum2024challenging}, anti-fascist~\cite{mcquillan2022anti}, abolitionist~\cite{benjamin2019race, barabas2020beyond, earl2021towards, jones2021we, williams2023no}, post-capitalistic~\cite{feltwell2018grand, browne2022future}, and anarchist~\cite{keyes2019human, linehan2014never, asad2017creating} thought.
Values-driven research have been put to practice, \eg, application of feminist HCI~\cite{dimond2012feminist, dimond2013hollaback}.
These strands of research are tied by their shared rejection of technological determinism~\cite{heder2021ai} in favor of recognizing that technology and society mutually shape each other.
It calls for imagining alternative liberatory futures, aptly summarized by \citet{keyes2019human} as: ``radically reorienting the field towards creating prefigurative counterpower---systems and spaces that exemplify the world we wish to see, as we go about building the revolution in increment''.

In IA, critical information theory~\cite{fuchs2009towards} calls for the study of how information and IA relates to the ``processes of oppression, exploitation, and domination'' and to conduct all research activities in ``solidarity with the dominated and for the abolishment of domination''.
Echoing similar sentiments, emancipatory IA~\cite{mitra2025emancipatory} is defined\footnote{
While the original definition was in relation to information \emph{retrieval}, we extend the scope to information \emph{access} as a whole.
} as ``\emph{the study and development of information access methods that challenge all forms of human oppression, and situates its activities within broader collective emancipatory praxis}''.
Emancipatory IA aspires for universal humanization and the elimination of all structural forms of oppression including colonialism, racism, patriarchy, casteism, transphobia, religious persecution, and ableism.
Similar to other values-driven research, emancipatory IA rejects the premise of technological determinism in favor of recognizing the agency and responsibilities of researchers and developers to affect positive societal transformations via IA technologies.
The framework proposes a nonexhaustive list of projects including: safeguarding against capture, surveillance, manipulation, dehumanization, and inequitable outcomes, and promoting emancipatory pedagogy.
While emancipatory IA specifies \textit{what} projects we should embark on, we theorize \textit{how} we should work on these projects.
Specifically, Freirean design entrusts communities to come up with new imaginaries of IA systems and actualize them.
In doing so, it challenges the technologists-as-liberator frame and articulates the need for intentional social arrangements in whose context emancipatory IA's theories can be put into practice.

Incorporating liberatory epistemologies is critical to the pursuit of social justice and emancipation in computing.
However, we should not valorize ``universal methods''~\cite{avle2017methods}.
Structures of oppression manifest differently for different marginalized communities and should be interpreted in light of their historical relations.
This motivates the need for community participation in the making (and unmaking) of technologies as we discuss next.

\subsection{Community participation in technology production}
\label{sec:background-justice}
Central to our framework is the notion that technology cannot truly be emancipatory without the reflective participation of communities\footnote{In this context, by `community' we refer to groups of people related by shared intersectional identities.}.
Existing models for collaboration with wider communities---such as participatory design (PD)~\cite{bvarphidker1995cooperative, sanders2008co, qi2025participatory, floyd1989out}, participatory action design research (PADR)~\cite{bilandzic2011towards, haj2016padre}, and metadesign~\cite{fischer2000meta, fischer2003meta}---have been applied in HCI.
We make explicit how Freirean design diverges from these existing models of community participation and argue that these approaches maintain centralized power and responsibility that reinforces the dependency of communities on technologists to meet their needs.

PD originated from the Scandinavian ``Cooperative Design'' movement during the '60s and '70s, and emphasizes collaboration between technologists (or designers) and relevant stakeholders (often end-users) such that the final design outcome reflects and incorporates the opinions and desires of all stakeholders.
PD is inherently visionary; it aims for technologists and users to cooperatively envision and build idealized systems, saliently making a political commitment to involve users in design-decisions that affect them.
In practice, PD has been leveraged in the development of initiatives across multiple fields including tech and computing~\citep{duarte2018participatory, sabiescu2014emerging}, social programs and community projects~\citep{asad2017creating, creativereactionlab2018}, and urban and environmental design~\citep{baumann2017infrastructures}.

PADR combines PD with Participatory Action Research (PAR)~\cite{bradbury2007sage}.
PAR emphasizes collaboration between researchers and community members engaged in collective inquiry.
In PAR, ``communities of inquiry and action evolve and address questions and issues that are significant for those who participate as co-researchers''~\cite{bradbury2007sage}.
PAR, in turn, draws from Action Research (AR)~\cite{hayes2011relationship, rapoport33197, lewin1946action, baskerville1999investigating} that parallels the Freirean literacy movements~\cite{campos2022introduction}.
By combining PAR with PD, PADR invites community participation beyond the design phase; it adopts an iterative approach for community collaboration for greater collective understanding of the problem to affect desired social change.

Of the different existing participatory frameworks, metadesign is the closest to our conceptualization of Freirean design.
Metadesign extends the collaboration with users ``to include an ongoing process in which stakeholders become co-designers---not only at design time, but throughout the whole existence of the system''~\cite{fischer2003meta}.
To facilitate this, it recommends that systems be \textit{underdesigned} at design time allowing users to create appropriate solutions for themselves, and to treat `designing the design process' as a critical activity.
Metadesign and PADR contrast with PD in that PD typically has an end-point when the design is fixed and can no longer be changed by users~\cite{fuad2013design}.

\citet{design2020costanza-chock} highlights several pitfalls of PD to explain the continual de-prioritization of marginalized voices.
Common shortcomings include selected participants being unrepresentative of all potential users, or resembling the researchers themselves, and economic pressures to appeal to the most profitable users.
Additionally, naive implementations of PD---``Add Diverse Stakeholders and Stir''~\citep{delgado2021}---can lead to de-politicizing of well-intentioned participation, whereby a focus on individual stakeholder needs obscures the consideration of power structures and collective good ~\citep{delgado2023}.
Seemingly inclusive and democratic enactments for participation can reinforce existing power structures ~\citep{design2020costanza-chock}, reduce participation to consultation without affording users real power~\cite{corbett2023power}, allow for participation-washing~\cite{sloane2022participation}, and result in co-option of community knowledge and labor~\cite{birhane2022power}.
Such outcomes are best summarized: ``je participe, tu participes, il participe, nous participons,
vous participez, ils profitent'' (in English: ``I participate, you participate, he participates, we participate, you participate, they profit'').\footnote{This quote is from a French poster by Atelier Populaire (May 1968). V\&A Museum collections accession number E.784-2003.}

Realized examples of these frameworks still tend towards centralized power and the primacy of technologists, since technologists maintain executive authority to decide the parameters of participation~\citep{delgado2023} and have ultimate power to block stakeholder proposals in favor of their own ideas~\citep{delgado2021}.
Even in metadesign, which aims to empower users to be co-designers, this distinction remains, evident from the following quote~\citep{fischer2003meta}.
\begin{quoting}
  ``\emph{The goal of making systems modifiable by users does not imply transferring the responsibility of good system design to the user. In general, ``normal'' users do not build tools of the quality a professional designer would since users are not concerned with the tool per se, but in doing their work.}''
\end{quoting}

The tension here emerges from a critical difference in the definition of ``users'' operationalized in our work in contrast to previous frameworks.
While in previous framings a user is defined by their lack of design know-how and system development expertise, we distinguish technologists and users based on who holds power to change the technology and who needs to seek external approval.
The ``user'' in our framework are marginalized communities who oftentimes hold relevant expertise that are rendered invisible by the WEIRD (Western, Educated, Industrialized, Rich, and Democratic) gaze.
Examples like the Māori Data
Governance Model~\cite{kukutai2023maori} and the rich history of technological co-option in the Brazilian favelas~\cite{nemer2022technology} challenge this dominative imaginary of unskilled masses of marginalized peoples who must be saved from the harms of technology by altruistic technologists.
Instead, our analysis is better served by identifying the structural barriers preventing communities from building for themselves, such as the lack of resources and legitimate access to modify system internals.
Such radical proposals for community participation in shaping technologies cannot succeed in a silo, and should begin by connecting the dots between ongoing tech workers movements for platform cooperativism~\cite{scholz2015think, schneider2018internet, scholz2016ours, scholz2017uberworked}, solidarity economy initiatives~\cite{bergeron2015social, egorov2021cooperation}, and unionization drives~\cite{mathenge2024data, greenhouse2021amazon, marshall2021google}.
The removal of central authority further creates a need for decentralized governance with a mandate for emancipatory outcomes.
We foreground these requirements in Sections~\ref{sec:freire} and~\ref{sec:problem}.

\begin{figure}
    \centering
    \includegraphics[width=\columnwidth]{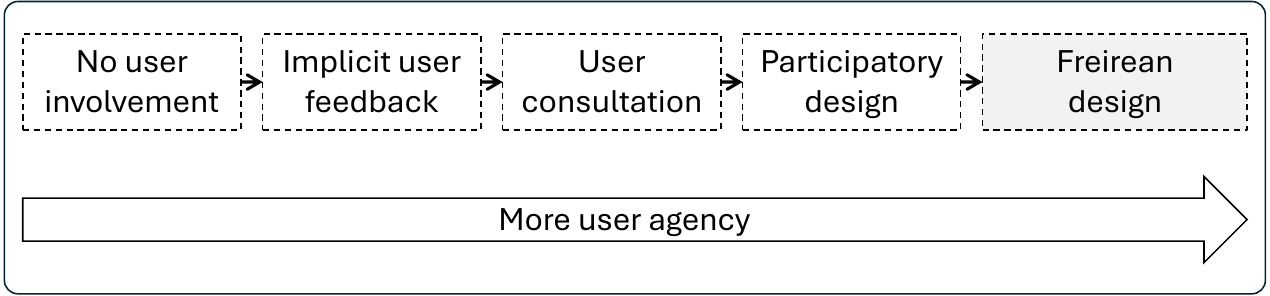}
    \caption{The spectrum of user participation in the different paradigms for IA technology  development.}
    \label{fig:freirean-design}
\end{figure}

In IA, active participation of users are largely lacking in the context of proprietary platforms owned by big corporations.
It is commonplace for these platforms to employ their own interpretations of what users want based on extensive surveillance and modeling of their behavioral signals---\eg, clicks~\citep{jiang2013mining, chuklin2015click} and mouse movements~\cite{diaz2013robust}---potentially complemented by limited user consultations.
There has been recent calls to move beyond this behaviorism paradigm~\cite{ekstrand2016behaviorism} towards a culture of co-constructing the platform with users using PD methodologies~\cite{ekstrand2025recommending}.
Our work goes further, to move towards not just building with users, but enabling them to build for themselves, as visualized in Figure~\ref{fig:freirean-design}.

\subsection{Resisting capture}
\label{sec:background-capture}

\paragraph{Decentralized social media. }
In the social media landscape, there have been longstanding concerns about how platforms' unequal application of moderation rules creates unfair outcomes~\citep{diaz2021double} and how platforms make decisions at the expense of their users~\citep{milmo2022instagram}.
These concerns expanded to include explicit ideological capture after Musk's takeover of Twitter in 2022~\citep{leerssen2025murdoch, haggart2022elon, mastodon2025people, bsky2025free}.
In response, protocol-based decentralized social media like the ActivityPub-based Mastodon\footnote{\url{https://www.w3.org/TR/2018/REC-activitypub-20180123/}} and ATProto-based Bluesky\footnote{\url{https://bsky.social/about/blog/10-18-2022-the-at-protocol}} emerged as alternatives that explicitly center resistance to elite and corporate capture, and promote individual and community agency, \eg, in feed curation and moderation.
The decentralized paradigm raises its own issues, including:
\begin{enumerate*}[label=(\roman*)]
    \item The degree to which nominally decentralized networks actually facilitate decentralized operation~\citep{Yu2025IdentifyingRI} and effectively avoid capture,
    \item the tension between democratic control of sections of federated networks and the access to a (relatively) unified discourse on a centralized platform \citep{fraser2025beyond}, and
    \item how decentralization of technological infrastructure alone tend to reproduce dominant structures of oppression~\citep{hendrix2022whiteness}.
\end{enumerate*}
However, decentralized social media remains under active development; their protocol-based nature structurally encouraging community development.
The ATProto ecoysytem, for example, now features independently developed interoperable services for blogging\footnote{\url{https://leaflet.pub}, \url{https://pckt.blog/}}, scientific preprints\footnote{\url{https://chive.pub/}}, photo sharing\footnote{\url{https://www.flashes.blue/}}, and source control\footnote{\url{https://tangled.org/}}.
Emancipatory IA should draw insights from these efforts and contribute back new insights and sociotechnical innovations.

\paragraph{Open Web Search. }
The Open Search Foundation\footnote{\url{https://opensearchfoundation.org/about-open-search-foundation/}} is engaged in projects that are particularly relevant to our work.
This includes the OpenWebSearch.EU project\footnote{\url{https://openwebsearch.eu/the-project/}} and the \#FreeWebSearch Charter~\citep{open2025freewebsearch}.
The OpenWebSearch.EU project began in response to growing concerns about the concentration of power over IA platforms in the hands of a few large corporations bound by national interests.
According to the project website: ``\emph{Despite being a backbone of our digital economy, web search is dominated and limited by a few gatekeepers like Google, Microsoft, Baidu or Yandex. [\dots]
This imbalance endangers democracy and limits the innovative potential of Europe’s research landscape and economy.}''
They further clarify their intention to contribute towards Europe’s digital sovereignty.

These Open Search projects and charters significantly overlap with calls for emancipatory IA and presents critical opportunity for collaborations across proponents of both missions.
However, there are also critical divergences in motivations and aspirations between the two perspectives.
The Open Search Foundation is focused on Europe’s digital sovereignty and is centered on European values, principles, legislation, and standards.
While, it represents important initiatives to build counter-structures to challenge US Big Tech's outsized dominance over global IA platforms, it does not speak for the values and needs of the global south that represents 85\% of the world's population\footnote{The global south's population estimate is based on~\citet{veron2023much}.}.
We should recognize that \emph{Eurocentrism} is not a meaningful antidote to \emph{American Exceptionalism}, and no IA platform is truly open nor reflects the internet in all its diversity if it systemically erases the voices of the global south.
This model of challenging US Big Tech reflects the values of multipolarism in which influence over our information ecosystems is concentrated in the hands of few powerful state and supernational actors within the international order (\eg, USA, EU, China, and Russia); at the cost of further disenfranchising other often-marginalized global populations.

In contrast, emancipatory IA seeks epistemic justice for both the global south as well as for marginalized populations within powerful nation-states, and is saliently decolonial in its aspirations.
These differences are not simply philosphical divergences, but have material consequences for platform design and research directions.
For example, the absence of trustworthy centralized institution(s) to steer platform maintenance and governance, exacerbates the need for decentralization in emancipatory IA.
Emancipatory IA also requires reflective participation of marginalized communities in ways that an open search platform for EU does not.
Further, emancipatory IA espouses different approaches to addressing societal concerns compared to OpenWebSearch.EU and the \#FreeWebSearch Charter.
For example, the \#FreeWebSearch Charter promotes transparency and diversity to tackle political bias and dominant narratives.
However, that risks encoding false equivalence between oppressive and emancipatory politics which may contribute to \emph{algorithmic bothsidism}~\citep{mitra2025search}.
Emancipatory IA, in contrast, pushes for structural ways to counter systemic power differentials and uplift marginalized voices.
That is, while the \#FreeWebSearch vision promotes what may be considered a liberal model of democracy, emancipatory IA aligns with a more critical model of democracy~\citep{vrijenhoek2021recommenders}.

However, these differences notwithstanding, we believe that both these approaches can benefit from one another.
The digital sovereignty framing is likely to attract significantly more funding and investments that may enable projects like OpenWebSearch.EU to prototype some the technologies that may turn out to be critical for emancipatory IA.
Emancipatory IA research may in turn unlock new innovations in the space of sociotechnical design of IA platforms that provides structural mechanisms to ensure compliance with the demands of \#FreeWebSearch.
\section{Information Access of the Oppressed}
\label{sec:freire}
``\emph{Pedagogy of the Oppressed}'' (original Portuguese title: ``\emph{Pedagogia do Oprimido}'')~\cite{freire1970pedagogy} authored by Brazilian Marxist educator Paulo Freire was first published in 1968.
The book presents a detailed Marxist class analysis exploring oppressor-oppressed relations in the context of pedagogy.
In his canonical work, Freire critiques the traditional ``banking model of education'' in which it is the task of the teacher to be the arbiter of what constitutes valid knowledge and to ``deposit'' their teachings, while it is the duty of the students to ``\emph{patiently receive, memorize, and repeat}''.
Instead, Freire advocates for a problem-posing dialog between teacher and student that engages the latter in critical reflection and sensemaking of their own state of oppression.
In Freire's approach, teachers and students co-construct knowledge and collectively engage in \emph{praxis}---\ie, ``\emph{reflection and action directed at the structures to be transformed}''.

While Freire's work centers on pedagogy and learning, we apply his theories to modern IA, where information seekers become `learners' via the search of digital information.
We argue that IA platforms are sites of liberatory struggle and should operationalize the Freirean approach by encouraging communities to co-opt and co-construct IA systems, and in turn shape how information is experienced, as part of critically examining their realities within oppressive structures, and taking action to transform them.
Specifically, we draw from Freire's theories to:
\begin{enumerate*}[label=(\roman*)]
    \item Surface the technologist-user dichotomy,
    \item challenge the technologists-as-liberator frame, and
    \item develop the notion of \textit{Freirean design} extending his problem-posing approach to IA.
\end{enumerate*}

\subsection{The technologist-user dichotomy}
Our work begins with the observation that we typically relate to technology either as technologists (who envision and build technology) or as users (who seek information via the platform).
Although technologists may sometime use the technologies they build, most users of technology do not typically contribute directly to their envisioning or development.
This technologist-user dichotomy mirrors the teacher-student relationship that Freire analyzes.
By delegating the tasks of envisioning, designing, and governing IA platforms exclusively to technologists, we in turn elevate them to the position of gatekeepers of knowledge and hand them tremendous influence over our collective sensemaking of our place and relationships in this world; while the rest of the population is designated to be the \emph{objects} of technological influence---their existence digitized for monitoring, modeling, and manipulation.
As Freire writes:
\begin{quoting}
  ``\emph{More and more, the oppressors are using science and technology as unquestionably powerful instruments for their purpose: the maintenance of the oppressive order through manipulation and repression.
  The oppressed, as objects, as `things,' have no purposes except those their oppressors prescribe for them.}'' (Page 60)
\end{quoting}

This designation of information seekers as agency-free ``\emph{objects}'' normalizes platforms  monitoring their online activities, including their clicks~\cite{chuklin2015click} and mouse movements~\cite{diaz2013robust} to farm signals for evaluation and training data.
The reduction of individuals to their interactions with the system incentivizes their manipulation even by non-platform actors---\eg, the growing practices of clickbaiting~\cite{donovan2014what} and rage-baiting~\cite{hom2015rage} by content producers.
Furthermore, this frame restricts our imagination to only be able to conceive the same technologists---or a subset of their altruistic converts---as those that should bear the burden of mitigating the risks to society and marginalized populations.

\subsection{Challenging the technologists-as-liberator frame}
Our dominant imaginaries of technology development frame the technologists as the sole executors of technological change.
Seeing through this frame one comes to the inevitable (but incorrect) conclusion that all research aiming to mitigate societal risks should focus solely on inventing mechanisms that IA platforms can operationalize to address concerns of bias, fairness, misinformation and disinformation, privacy, and safety; and we should ensure platforms employ said mechanisms by appealing to their better nature, through public pressure, or via enforcement of legal regulations.
But what remains elusive just beyond the reach of this frame is the idea that may be the control over these platforms should never have been centralized in the hands of large corporations, because that threatens Big Tech's grand project of empire building~\cite{hao2025empire}.
To those well-intentioned ``converts'', we quote the following text by Freire:
\begin{quoting}
  ``\emph{Our converts, on the other hand, truly desire to transform the unjust order; but because of their background they believe that they must be the executors of the transformation.
  They talk about the people, but they do not trust them; and trusting the people is the indispensable precondition for revolutionary change.
  A real humanist can be identified more by his trust in the people, which engages him in their struggle, than by a thousand actions in their favor without that trust.}'' (Page 60)
\end{quoting}

This is not an argument against research (or researchers) that develop mitigative intervention mechanisms, but rather a call for us to recognize that those efforts should be grounded in broader collective acts of movement-building to challenge the concentration of power in the hands of large multinational platforms and to ultimately \emph{abolish Big Tech}.
Otherwise, any positive impact of these efforts will be temporary and potentially contribute to \emph{ethics washing}~\cite{wagner2018ethics, van2022ai} and further entrenchment of corporate hold over our IA platforms.
Instead, we advocate that societally-motivated technologists should join the struggle in solidarity with the people who are most at risk and harmed by these technologies.
Dismantling the technologists-as-liberator frame will require radical rethinking of our approach to technology design, and we operationalize Freire's approach next to put forward such a proposal.

\subsection{Freirean design}
We define Freirean design as \textit{a design methodology and a political commitment to engage users in the active co-option and co-construction of technologies}.
The inclusion of those at the cross hairs of technology is much more than symbolic.
It is an act that recognizes that epistemic and representational justice cannot be ``gifted'' by technologists to marginalized communities.
It is also driven by the realization that the historical and cultural context of marginalization and the structures of oppression are not universal---and differ radically, while sharing some commonalities, from the inner city neighborhoods in the US to the favelas of Rio to the bastis of Kolkata.
It is only through community participation that we can truly aim for the necessary social, political, and cultural sensitivity in our pursuit for universal emancipation.
We should also be pragmatic that technologists---many of who are employed in the for-profit technology industry or depend on research funding from the military-industry complex---are unlikely to take the necessary risks to execute revolutionary change to challenge structural oppression faced by others `out there'.
But for community members, their participation becomes a radical act of connecting their existential struggles against oppression with the making and unmaking of technology, and in turn their participation becomes a site for solidarity and movement building.

When designing new modes of participation to center the needs of the marginalized, we should counter cynicism about the likelihood of broad public engagement by reminding ourselves that communities have long been resisting oppression and appropriating everyday technologies to alleviate their oppression~\cite{nemer2022technology}.
It is true that the ability to appropriate and rework technology is considerably more limited in the age of online services, since the ``code'' and infrastructure is not easily accessible to or modifiable by users, relative to the ease of appropriating physical devices.
Therefore, we cannot assume a reluctance to engage, but should refocus our energies in identifying structural barriers to their participation and rethink online technology stacks to re-expose the cracks that can be transformed into legitimate spaces for not just participation, but appropriation in support of everyday struggles against oppression.
These spaces should be structurally guaranteed in our design to support people marginalized by these technologies to reflect, negotiate, and collaborate with each other to co-opt and co-construct the technology to suit their own needs.

To re-construct technology in support of emancipation, we should unequivocally reject the premise of technological determinism~\cite{chandler1995technological} and confront the repeating cycles of technodeterministic hype coming from our own technology communities---\eg, just in recent years: Blockchain, Non-fungible tokens (NFTs), Virtual Reality, and now AI.
This is not to deny that many of these technologies have their material usefulness,\footnote{OK, may be not NFTs.~\cite{yang2023vast}} but the ask is to recognize that each of them drove a frenzied hype that tried to convince the world that the ubiquity of the technology was inevitable.
Consequently, we should also break the ever-repeating cycles of critiquing emerging technologies and pursuing insufficient means of mitigation when our resistance is better directed at those who wield power over these technologies to feed their personal need for wealth and power.
Instead of only obsessing about mitigations, we should build for something better, something for our collective selves, something joyful and emancipatory.
To quote Freire:

\begin{quoting}
  ``\emph{In order for the oppressed to be able to wage the struggle for their liberation, they must perceive the reality of oppression not as a closed world from which there is no exit, but as a limiting situation which they can transform.
  This perception is a necessary but not a sufficient condition for liberation; it must become the motivating force for liberating action.}'' (Page 49)
\end{quoting}

So, in this new frame, the role of the technologist becomes two-fold.
They should present the material risks of emerging technologies for the marginalized as an object of study to the people (users), mirroring Freire's problem-posing approach, to encourage them to see themselves as the necessary makers and unmakers of technology in support of their liberation.
Simultaneously, technologists should re-envision the structure of our online technology stacks---expanding on the ethos of `underdesigning' systems at design time and putting critical emphasis on `designing the design process' as popularized by proponents of the metadesign movement---such that they facilitate marginalized communities co-opting and co-constructing the technology to build pedagogical information experiences that encourage information seekers to critically examine their relations with oppressive structures, and to take action in support of their emancipation.
As the instigators of this work---and as recovering ``converts'' ourselves who previously approached the sociotechnical concerns of IA through the traditional lenses of fairness, transparency, and related frames---we see ourselves in solidarity with the marginalized engaged together in the struggle to free our IA platforms from authoritarian control and to rid ourselves from our complicity in the global rise in authoritarianism.
And in our new role, we recognize the importance of the practices of problem-posing in the process of affecting social and technological transformation.
\section{Putting Freirean Design to practice}
\label{sec:problem}
Freirean design challenges those in the traditional roles of technology design and development to open up the design process for community co-option and to dismantle centralized ownership in favor of public control.
But it is not simply about a transfer of ownership; Freire clearly articulates the need for problem-posing dialog for political consciousness raising as a vital element of the process to ensure that the ultimate outcome is emancipatory.
Likewise, in our proposal for Freirean design, we view problem-posing for consciousness-raising as a critical design practice that should encourage community members to critically reflect on the impact of technology on power relations between dominant and oppressed groups.
The starting point for putting Freirean design to practice therefore involves identifying design problems that engages communities in political reflection and liberatory actions.

In the context of IA, we may group these problems based on whether they relate to the design of platforms or information experiences.
The problems of \textit{platform design} concerns with safeguarding against capture, strategic underdesigning of the platform to make space for community co-construction, and developing democratic frameworks for governance; and the problems of \textit{information experience design} focuses on how we can experience information that helps us realize epistemic and representational justice for intersectional communities---\eg, pedagogical experiences on queer Dalit history.
In this section, we identify a seed set of problems for platform design.
Problem-posing for information experience design should follow in parallel, but should be initiated by community members in context of their own struggles, and is out of scope of our current work.

We choose the problems for platform design to prompt critical reflection, ideation, prototyping, experimentation, and studies.
We intend for each of these problems to solicit a diverse range of sociotechnical design ideas that foregrounds our emancipatory aspirations.
Each problem may be considered in isolation, but ultimately we seek systems that can jointly address these concerns.
Through these problems, we aim to expose a rich design space of alternative approaches that communities can explore to challenge the status quo.
We group these platform design problems into two themes:
\begin{enumerate*}[label=(\roman*)]
    \item Platform operations and governance and 
    \item safeguards from authoritarian capture and capitalist co-option.
\end{enumerate*}

\subsection{Theme: Platform operation and governance}
The problems posed under this theme aim to provoke rethinking of the design of our technological stack to create the conditions for community co-option and co-construction of IA technologies.
This includes the need to explore democratic governance frameworks, effective policy enforcement mechanisms, and research and tooling support for downstream information experience design.

\begin{enumerate}[label=\textbf{Prob\arabic*.}, labelindent=0pt, itemindent=3em, leftmargin=*]
    \setcounter{enumi}{0}
    \item \emph{How do we design the platform from the margins to legitimize community co-option and co-construction?}
\end{enumerate}
The platform should legitimize its own co-option and co-construction through  strategic underdesigning.
Community members should be empowered to build new IA artifacts---\eg, information retrieval components, moderation services, and information experiences---leveraging each other's artifacts in turn or building competing artifacts if existing ones are inadequate or deemed harmful.
These interactions in the design process should facilitate intra- and inter-community negotiations.
In this context, the platform should ensure an equal playing-field for community members; participation in co-construction should not be biased by who has access to more resources and labor~\cite{neophytou2026open}.
Models of engaging communities should be informed by theories of power and strive for epistemic justice, and to ensure that we do not extract free labor from already marginalized communities, it should consider mechanisms for their compensation and long-term sustainability of this work.

\begin{enumerate}[label=\textbf{Prob\arabic*.}, labelindent=0pt, itemindent=3em, leftmargin=*]
    \setcounter{enumi}{1}
    \item \emph{How do we develop a critical democratic governance framework?}
\end{enumerate}
Community participation is not a miracle-cure for all societal concerns with IA.
The opportunity to co-opt may be leveraged by oppressors as well as the oppressed.
We see these tensions already in context of app stores where their ownership by large for-profit corporations often imply that applications which harm marginalized communities are allowed to exist~\cite{molla2019rise} while those that oppose authoritarian violence are pulled down under government pressure~\cite{allyn2025legal}.
Addressing these with appropriate cultural and sociopolitical nuance requires community participation within robust democratic governance frameworks.

All platforms moderate~\cite{gillespie2018custodians}.
Platform governance need to deal with critical moderation policies and mediate challenging cases, \eg, the case of the `Napalm girl' photo~\cite{levin2016facebook}.
Content moderation and governance are not at odds with emancipatory aspirations; mitigating real-world harms is a core tenet of emancipatory praxis.
The challenge is to develop frameworks for democratic decision-making that are robust to capture.

\begin{enumerate}[label=\textbf{Prob\arabic*.}, labelindent=0pt, itemindent=3em, leftmargin=*]
    \setcounter{enumi}{2}
    \item \emph{How do we develop robust enforcement mechanisms for platform policies?}
\end{enumerate}
Democratic decision-making should be complemented by robust enforcement mechanisms.
Our aspirations for community co-option and co-construction is not a loophole for circumventing moderation nor an excuse for harmful delays in mitigative actions---\eg, to ensure removal of child sexual abuse material (CSAM) from circulation.
However, there is a natural tension here between developing infrastructure for the enforcement of policies and recreating infrastructure of censorship and centralized ideological control.

\begin{enumerate}[label=\textbf{Prob\arabic*.}, labelindent=0pt, itemindent=3em, leftmargin=*]
    \setcounter{enumi}{3}
    \item \emph{How do we support information experience design to promote emancipatory pedagogy?}
\end{enumerate}
Emancipatory IA challenges us to reimagine information experiences that are pedagogical and foreground critical scholarship.
In the context of platform design, this motivates fundamental research and development of shared tools for communities to design information experiences in service of their own emancipation.
Pedagogy in this context should not replicate the ``banking concept of education''---nor act as systems for indoctrination---but encourage reflection, critical thinking, and dialog.
This should be informed by multidisciplinary perspectives---including HCI, design, critical theory, social and political sciences, humanities, and arts---and search-as-learning research~\cite{urgo2025search}.

\subsection{Theme: Safeguards from authoritarian capture and capitalist co-option}
No liberatory struggle has the luxury to step outside of its own cage to architect a new liberated reality; the work begins from within the bounds of oppressive systems to transform the state of oppression to one of emancipation. 
The work on emancipatory IA have to likewise begin within present-day structures of oppression.
This motivates our work to prioritize safeguarding the platform from capture, blocking, political and economic coercion, and capitalist co-option.

\begin{enumerate}[label=\textbf{Prob\arabic*.}, labelindent=0pt, itemindent=3em, leftmargin=*]
    \setcounter{enumi}{4}
    \item \emph{How do we safeguard the IA platform against authoritarian capture?}
\end{enumerate}
The platform should ensure that it cannot be pressured to take an action (or be blocked from taking an action) via state regulations or orders, through corporate or other institutional mandates, using economic or other incentives, by threats or acts of force or violence, or other forms of coercion.
This requires that the infrastructure and its management, the governance of the platform, its sources of funding, and other aspects be decentralized to avoid exposing weak points that oppressive-actors can put pressure on to capture the platform.
Any serious attempt at building an emancipatory IA platform should strategically consider adversarial forms of platform capture and develop necessary mitigation.

\begin{enumerate}[label=\textbf{Prob\arabic*.}, labelindent=0pt, itemindent=3em, leftmargin=*]
    \setcounter{enumi}{5}
    \item \emph{How do we ensure that access to the platform cannot be blocked by authoritarian forces?}
\end{enumerate}
The platform should support reasonably convenient mechanisms to circumvent attempts by governments or Internet service providers to block access to the service.
There are precedents for online platforms, \eg, the messaging app Signal~\cite{whittaker2024proxy} that provide mechanisms for circumventing authoritarian block on the app.
Emancipatory IA should support similar mechanisms to ensure continued access to information under threats of authoritarian censorship.

\begin{enumerate}[label=\textbf{Prob\arabic*.}, labelindent=0pt, itemindent=3em, leftmargin=*]
    \setcounter{enumi}{6}
    \item \emph{How do we make the platform sustainable long-term in the face of political and economic pressure?}
\end{enumerate}
Any platform that succeeds in, or attempts to, aid liberatory struggles would likely come under immediate political pressure.
Beyond safeguarding against capture, the project of emancipatory IA should also consider how to ensure the physical and economic safety of those conducting the research and actively participating in development of the platform, including the community members.
Safeguards may range from funding support and mutual aid, legal and regulatory protections, community norms, technological safeguards, and other mechanisms.
Emancipatory IA platforms would also likely face severe economic competition from capitalist enterprises, and corporations might be particularly motivated to make the economics of the emancipatory IA platform untenable for its survival.
Therefore, the economics of the platform should be a critical focus of the project. 

\begin{enumerate}[label=\textbf{Prob\arabic*.}, labelindent=0pt, itemindent=3em, leftmargin=*]
    \setcounter{enumi}{7}
    \item \emph{How do we safeguard the platform from capitalist co-option and extraction?}
\end{enumerate}
Emancipatory IA platforms risk being co-opted by adversarial actors.
For-profit platforms, that do not share our emancipatory values, may attempt to extract economic value from the emancipatory IA platform without reciprocity.
The platform design should be intentional in safeguarding against these risks and codify the need for reciprocity through technological design, and licensing and other legal means---\eg, Copyleft licensing~\cite{frantsvog2012all}.
Emancipatory IA should draw insights from other justice-oriented initiatives, such as the free software movement~\cite{stallman1998why}.
\section{The Salient Tensions of Emancipatory IA}
\label{sec:discussion}
Emancipatory IA situates itself within the conflict between oppression and liberation; and resolves that tension by making a normative commitment to side with the oppressed.
Attempting to simultaneously address the many problems enumerated in Section \ref{sec:problem} will likely require us to negotiate other critical tensions---spanning the technological, social, political, economic, epistemic, legal, and personal.
We briefly discuss a short selection of such tensions in this section.

\paragraph{Community participation vs. Agility}
In the context of Freirean design, there may be legitimate concerns about the feasibility of community participation at scale and its impact on the agility of platform development.
While, such concerns may mirror known challenges in scaling PD~\cite{zahlsen2022challenges}, it is crucial to recognize the structural differences with Freirean design that may result in different outcomes.
Unlike in PD, in Freirean design there are no centralized group of technologists coordinating with communities.
So, while it is difficult to predict how Freirean design may scale with the number of participating communities without actually operationalizing Freirean design, we speculate that it will depend on the structures of communication \textit{with} and \textit{between} communities facilitated by the platform design.
Our argument is similar to that of Raymond's in ``The cathedral and the bazaar''~\cite{raymond1999cathedral}, where he counters Brooks's law (``Adding manpower to a late software project makes it later''), arguing that the structure of coordination within the project determine how the complexity and communication cost grow with project members.
Speculatively, the structure of coordination in Freirean design may look as follows: Communities develop information experiences independently of each other; others may combine these different information experiences, without explicitly seeking permissions, to create richer compositions; while communities retain the right to ``remove'' their information experiences from specific compositions.
Such distributed coordination may reflect ``the bazaar'' of information experiences, in contrast to the proprietary cathedrals of Google and Bing.
Done right, we posit this would meet community needs better than centralized platforms.

\paragraph{Platform vs. Protocol}
In response to the growing concentration of power of a handful of platforms over public speech and IA, there is an emerging debate over the platform \vs protocol approach~\cite{masnick2019protocols}.
This is particularly dominant in the social media discourse where protocols like ActivityPub and ATProto have gained popularity as alternatives to billionaire-owned platforms.
In this paper, we have liberally used the term ``platform'' to refer to the object of our envisioning in the context of emancipatory IA.
This is intentional, partly to sidestep this debate earlier in the paper and use ``platform'' as a shorthand to refer to whatever solution that emerges as appropriate; but also to reflect the realities of building for users who may not understand nor care about the mechanics of platforms \vs protocols. 
While protocols present distinct advantages over platforms for user ownership of data, competition, and resistance to single points of capture, as \citet{hendrix2022whiteness} argue, protocols in lieu of platforms do not inherently challenge the hierarchical social structures that they emerge from; 
the expected benefits of protocols often rest on a neoliberal belief that market competition will drive better outcomes.

\paragraph{Top-down vs. Bottom-up Design}
Design-related tensions emerge from opening up the platform to community co-option and co-construction.
When large masses of Twitter users abandoned the social media site in 2022 to flock to Mastodon, they experienced significant growing pains onboarding onto the new platform~\cite{diaz2022using}.
New users had to shoulder the burden to familiarize themselves with a new platform that posed significant friction, especially for non-tech-savvy users, and exercise trust-based actions (such as selecting a Mastodon instance) that many felt ill-equipped to do~\cite{stefanija2023solid}.
In enabling, co-option and co-construction, emancipatory IA should still ensure not to offload responsibilities to the user that might overwhelm them, and continue to prioritize the ease of usability of the platform and consistency of user experience.
It needs to drive a balance between the flexibility it affords to contributing communities and the overall usability of the platform.

\paragraph{Justice vs. Legality}
Lastly, we should be clear-eyed about challenges that emerge when the law diverges from justice.
We opened this article with examples of authoritarian attempts at capture using executive orders and mandates.
Under the conditions of oppression, the law disproportionately sides with the oppressor, which may create situations where the mandate of Emancipatory IA conflict with existing laws.
Emancipatory IA platforms need robust decision-making protocols to navigate these tensions while respecting legal frameworks (where possible) and prioritizing our stated commitments to human rights and social justice.
\section{Conclusion}
\label{sec:conclusion}
IA is too critical to our collective sensemaking and sociopolitical discourse to be left in the hands of powerful corporations and economic elites.
In a globalized world, our IA platforms should represent the interconnectedness of different communities and cultures worldwide, and not privilege the perspectives of a few powerful nations nor continue to reproduce the epistemic injustices of our past.
These injustices are the poison fruits of systemic oppression that have their roots in our history and continue their domination~\cite{collins1990black} into the present---\eg, colonialism, cisheteropatriarchy, classism, casteism, and ableism; and so it is the explicit goal of emancipatory IA to dismantle these structures.

Our approach to emancipatory IA is predicated on the rejection of technosolutionism.
It is not the design of the technology that confronts oppression, but rather it is the creation of spaces for community participation, collective deliberation, and democratic negotiation within the platform---that is so often missing in our modern technology design---that we believe holds the key to realizing emancipatory access to information.
The role of design is to structurally encode these spaces within the IA platform, which is necessary but in itself not sufficient for the platform to be emancipatory.
Emancipation after all is a political project and so it is the sociopolitical arrangements within these spaces that are key to determining whether the platform will be emancipatory or oppressive.
Emancipatory IA does not brush this political reality under the rug but rather foregrounds it.
To bend the arc of IA towards emancipation we need to experiment with different structures of democratic participation, governance, and decision-making.
But beyond structures and processes, we also need to ground these discourses and negotiations in critical scholarship, theories, and practices that distinguish structural understanding of systems of oppression from reactionary narratives.
For this, we should lean on relevant disciplines such as critical theory, library and information sciences, STS, and others; while being clear-eyed about the growing phenomenon of reactionary co-option and weaponization of social justice language to prop up false victimhood---\eg, \cite{strickland2016white, southernmens, savera2024haf}.
Our ``neutrality'' should manifest through equal treatment of all groups with respect to humanization and emancipation; not drawing of false equivalences between oppressors and oppressed.

Freirean design does not point to miracle-cures for the ailing growth of global authoritarianism and oppression.
Instead, it responds to the call for building an emancipatory movement within the IA community~\cite{mitra2025emancipatory} by articulating some of the key problems that envisioned emancipatory platforms should address.
Envisioning in this context should not be idle conjectures but \emph{praxis}; we should use these questions to reflect and act on the problems collectively and in the process co-construct the social arrangements of individuals and institutions committed to the causes of universal humanization and emancipation.

Our movement begins here.
We are moved not by the opportunity to imagine a different sociotechnical future but by the abject necessity of it in the face of growing authoritarianism and oppression.
We do not move alone but in solidarity with others because our liberation, and the struggle for it, are bound together.
We move not in fear but with courage and out of love for our own humanity and that of others.
Where the oppressor attempts to capture and control, we move to co-opt, co-construct, and liberate.
We move not just to negate oppression but to create something new, something joyful, something emancipatory.

\bibliography{references}
\end{document}